\documentclass{PoS}

\title{Measurement of the charged-pion polarisability at COMPASS}

\ShortTitle{Measurement of the charged-pion polarisability at COMPASS}

\author{\speaker{Alexey Guskov}%
        \thanks{on behalf of the COMPASS collaboration.}\\
       Joint Inst. for Nuclear Research (RU)\\
       E-mail: \email{alexey.guskov@cern.ch}}
       
\abstract{The electric (${\alpha}_{\pi}$) and the magnetic (${\beta}_{\pi}$) 
polarisabilities are fundamental properties of the pion
characterising the rigidity of
its internal structure.
 They have been precisely measured at the
COMPASS experiment at CERN with a ${\pi}^{-}$ beam of 190~GeV/c assuming
 ${\alpha}_{\pi}+{\beta}_{\pi}=0$. Muons of the same
momentum were used for controlling of systematic effects.
 The
obtained result ${\alpha}_{\pi}=-{\beta}_{\pi}=(2.0\pm 0.6_{stat.}\pm 
0.7_{syst.})\times 10^{-4} fm^3$ is in agreement with
the prediction of the Chiral Perturbation Theory.
}

\FullConference{The European Physical Society Conference on High Energy Physics\\
		22--29 July 2015\\
		Vienna, Austria}

\begin{document}

\section{Introduction}
In classical physics the polarisability of a medium or a composite system
is a well known properties related to the response of the system to the presence of
an external electromagnetic field. 
This concept can be extended to the case of hadrons - particles consisting of quarks bound by the strong interaction. This force s is strong compared to the electromagnetic force and the hadrons are accordingly compact and stiff if probed by the latter force. The electric polarisability $\alpha$ and the magnetic polarisability $\beta$ describe the rigidity of such objects against deformation by external electric and magnetic fields and can be tested in the reaction of Compton scattering off hadrons. They are the fundamental characteristics of hadrons and the comparison of theoretically predicted and directly measured values provides a stringent test of various theoretical models in the low energy region. For the proton, the polarisabilities $\alpha_p$ and $\beta_p$ were measured directly via Compton scattering off a hydrogen target decades ago. In contrast, for charged pions the experimental situation is more difficult since they are not available as fixed targets. Several attempts to measure the pion electric ($\alpha_{\pi}$) and magnetic ($\beta_{\pi}$) polarisabilities in dedicated experiments were performed. Groundbreaking work at Serpukhov used the process of pion radiative scattering off the nuclear target $\pi^- (A,Z) \rightarrow \pi^- (A,Z) \gamma$ \cite{serp1, serp2}, pion photoproduction $\gamma p \rightarrow \pi^+ n$ was used to determine the pion polarisabilities at Lebedev Institute \cite{lebe} and MAMI \cite{a2}. Some results were also obtained using the reaction $\gamma\gamma \rightarrow \pi^+\pi^-$ in $e^+e^-$ collisions \cite{pluto, dm1, dm2, mark2, Filkov, Kaloshin}. The experimental values for the quantity $\alpha_{\pi}-\beta_{\pi}$ assuming $\alpha_{\pi}+\beta_{\pi}=0$ are shown in Fig. 1. These obtained results are affected by large statistical and systematic uncertainties and their accuracy is a few times lower than the precision of theoretical predictions.  For instance, the Chiral Perturbation Theory (ChPT), the most successful model in the low energy region, predicts for the charged pion, the values $\alpha_{\pi}-\beta_{\pi} = (2.9\pm0.5)\times10^{-4}~fm^3$ and $\alpha_{\pi}+\beta_{\pi} = (-2.8\pm0.5)\times10^{-4}~fm^3$ \cite{Gasser}. So the new experimental input is needed.

\begin{figure}
\begin{center}
 \includegraphics[width=400px]{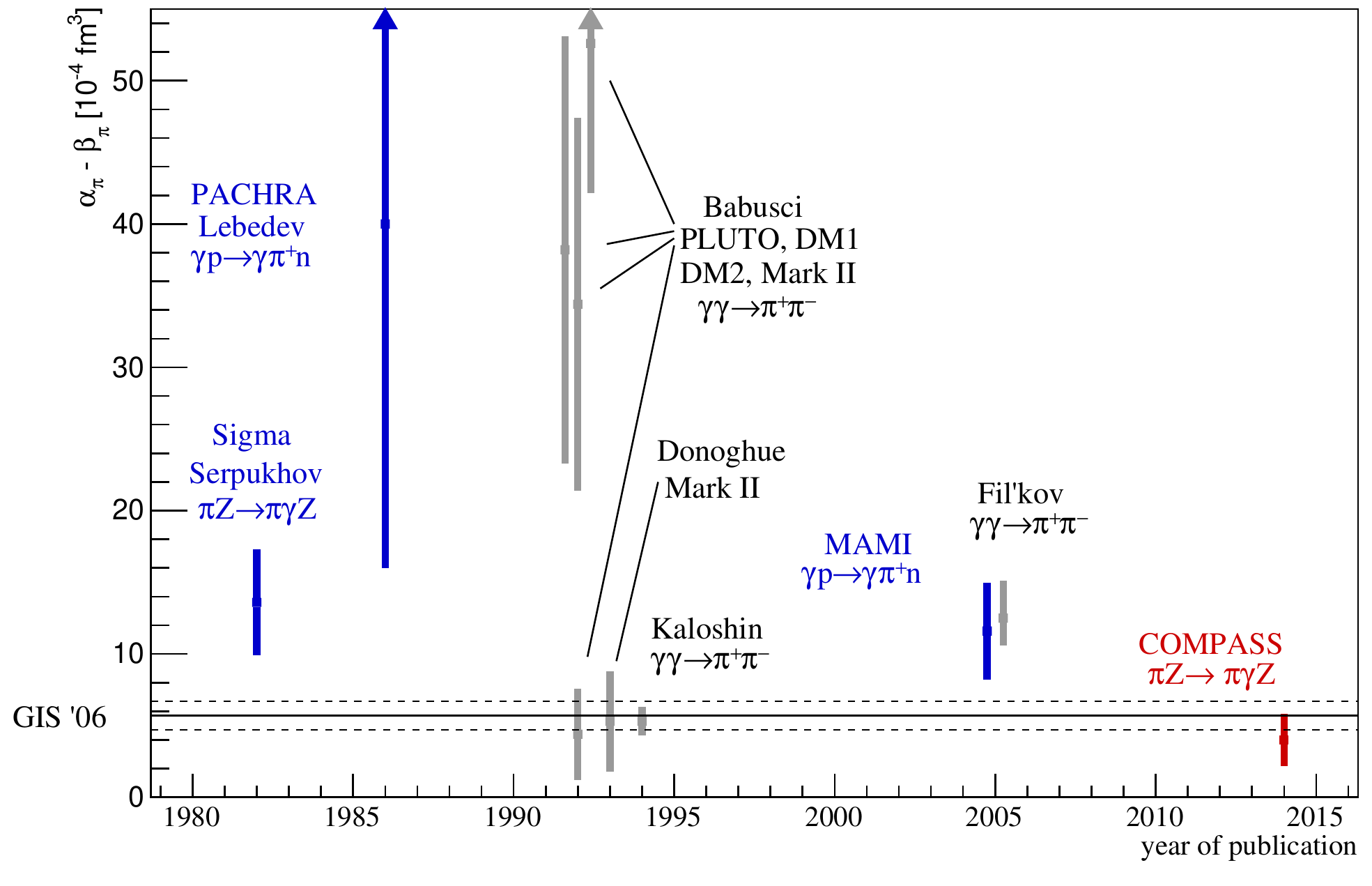}  
  \caption{\label{fig:diag}
The experimental results for the quantity $\alpha_{\pi}-\beta_{\pi}$ assuming $\alpha_{\pi}+\beta_{\pi}=0$. Horizontal lines represent the prediction of ChPT \cite{Gasser}.}
 \end{center}
\end{figure}

\section{Pion polarisability measurement at COMPASS}

\begin{figure}
\begin{minipage}{18pc}
   \includegraphics[width=200px]{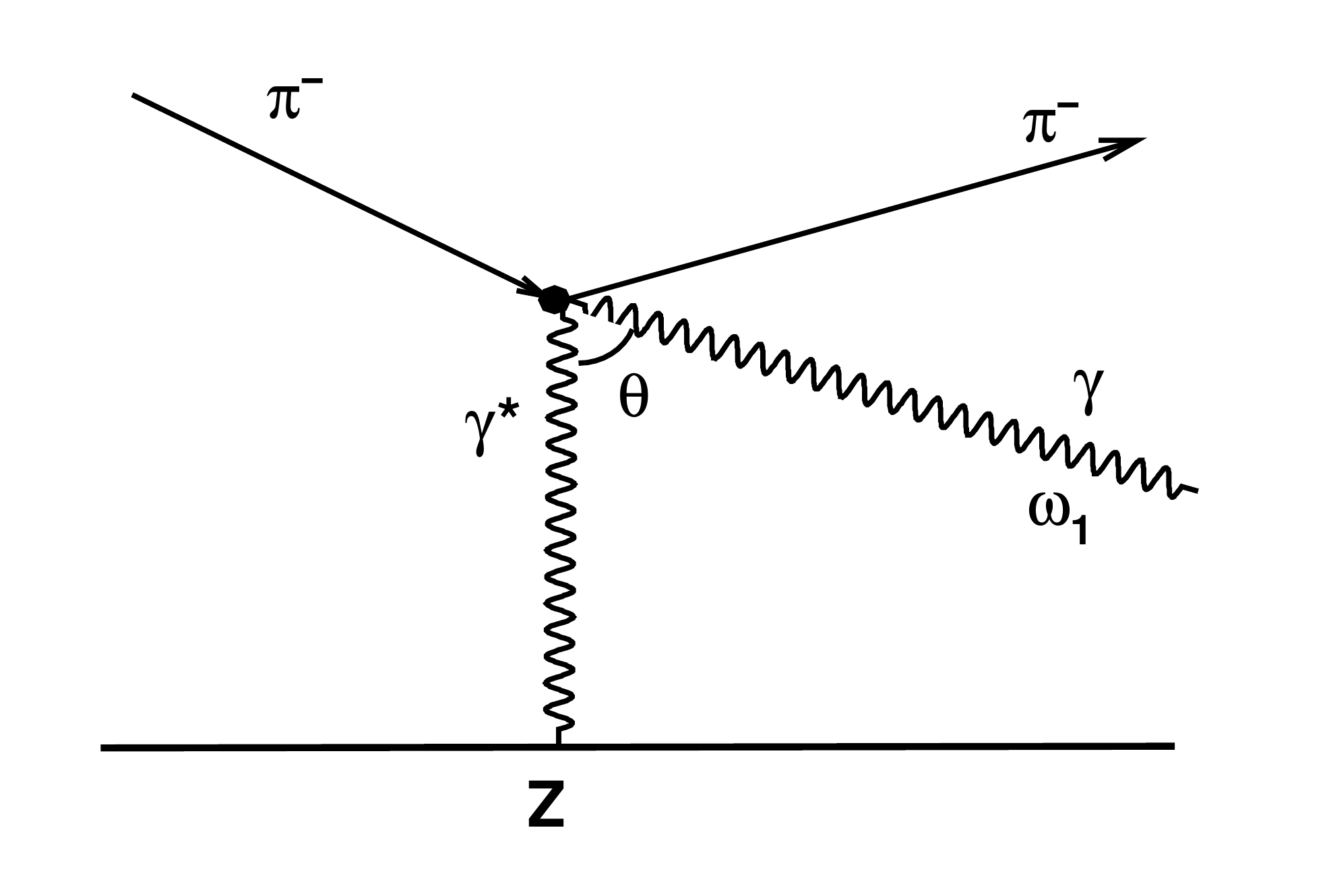}  
     \caption{\label{reac}Schematic diagram of the charged pion scattering off the nucleus with hard photon emission.}
\end{minipage}\hspace{2pc}%
\begin{minipage}{18pc}

  \includegraphics[width=200px]{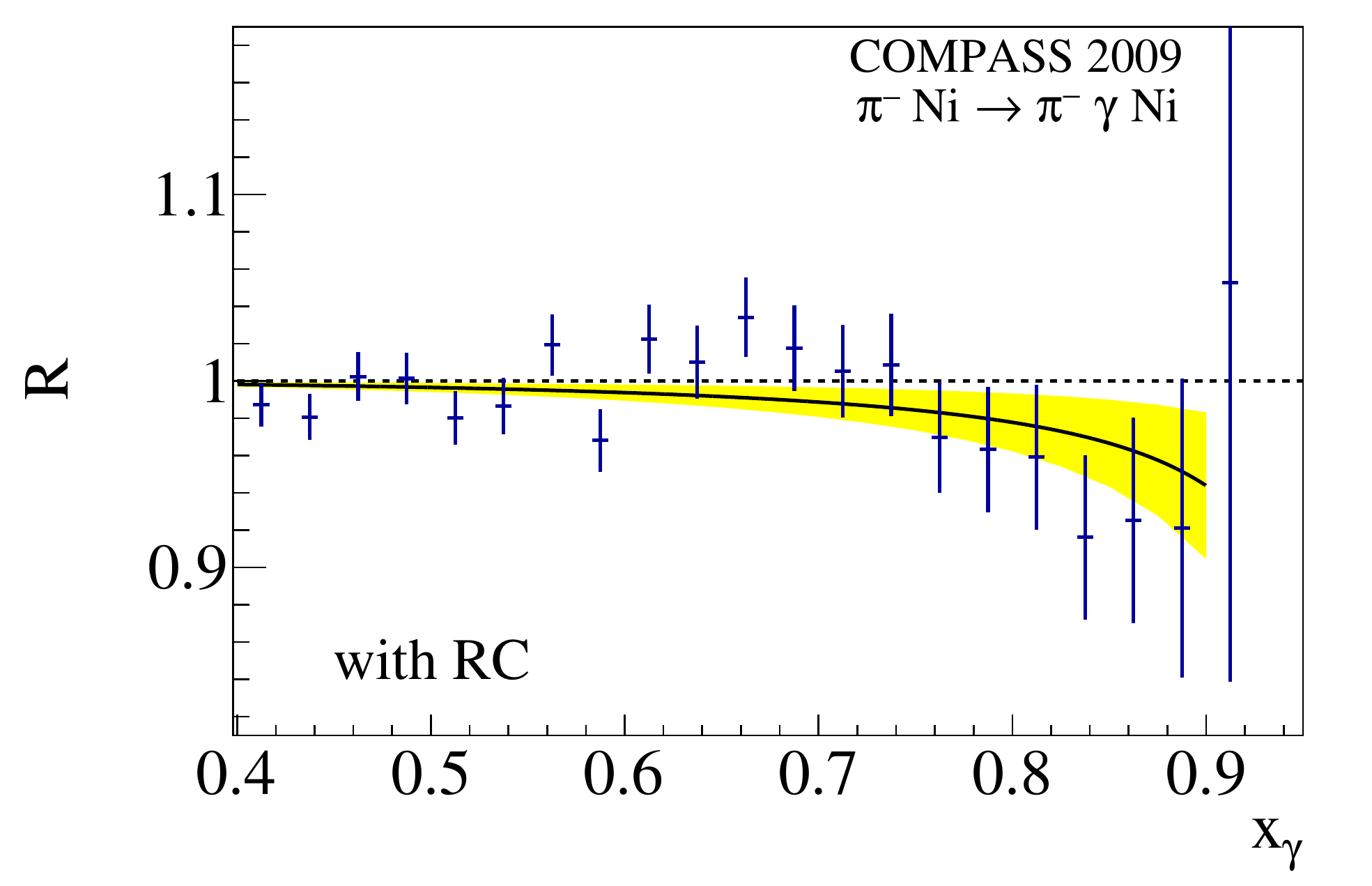}
    \caption{\label{fig:R} The measured ratio $R_{\pi}$ \cite{main}.}
 \end{minipage} 
\end{figure}

\begin{figure}
\begin{minipage}{18pc}
  \includegraphics[width=200px]{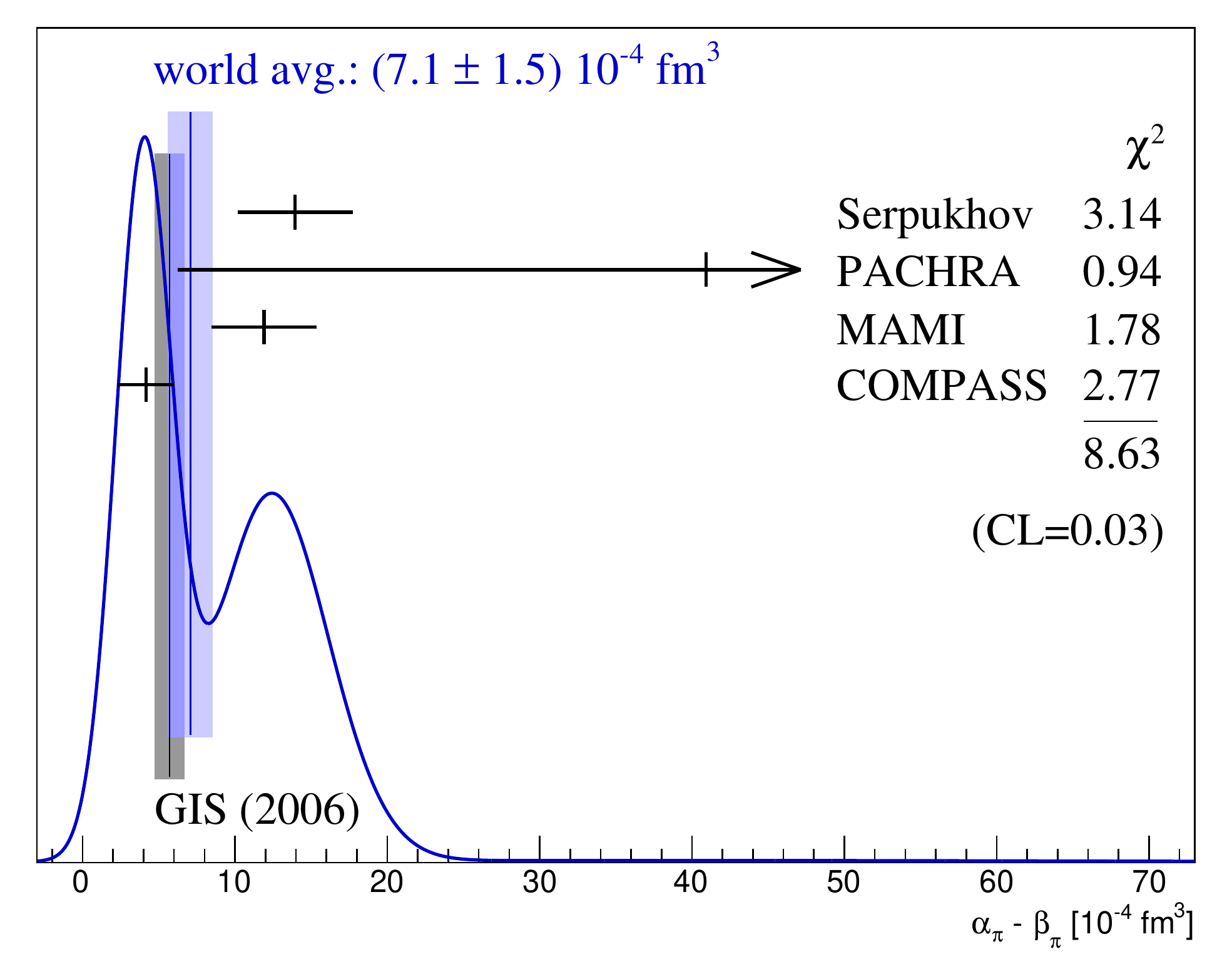}
    \caption{\label{fig:ideo} Ideogram for the value of $\alpha_{\pi}-\beta_{\pi}$ measured assuming $\alpha_{\pi}+\beta_{\pi}=0$ in the dedicated experiments including COMPASS.}
\end{minipage}\hspace{2pc}%
\begin{minipage}{18pc}

 \end{minipage} 
\end{figure}

COMPASS is a fixed target experiment at a secondary beam of the Super Proton Synchrotron at CERN \cite{proposal, proposal2, compass, compass2}. 
For the pion polarisability measurement at COMPASS the reaction $\pi^- (A,Z) \rightarrow \pi^- (A,Z) \gamma$,  which can be treated as Compton scattering of the pion off a virtual photon provided by the nucleus (see Fig. \ref{reac}), was used. Short data taking was performed in 2009. A hadron beam of 190 GeV/c scattered off the nickel target was used for the measurement. A muon beam with similar parameters was used to study  systematic effects. About 63 000 exclusive $\pi^-\gamma$ events in the kinematic range $p_{T}>40$ MeV/c, $m_{\pi\gamma}<3.5m_{\pi}$, $Q^{2}<1.5\times 10^{-3}$ GeV$^{2}/c^{2}$ and 0.4< $x_{\gamma}=E_{\gamma}/E_{beam}<0.9$ were used for pion polarizability extraction under assumption $\alpha_{\pi}+\beta_{\pi} = 0$. Here $m_{\pi\gamma}$ is the mass of the final $\pi\gamma$ state, $Q^2$ is the 4-momentum
transfer to the nucleus squared, $p_{T}$ is the transverse momentum of the scattered pion,  $E_{\gamma}$ and $E_{beam}$ are energies of the produced photon and the beam pion correspondently. The polarisability is determined from the $x_{\gamma}$ dependence of the ratio of the measured cross section $\sigma_{\pi \gamma}$ to the calculated cross section for the point-like pion
\begin{equation}        
  R_{\pi}=\left(\frac{d\sigma_{\pi \gamma}}{dx_\gamma}\right) \left/
  \left(\frac{d\sigma^{0}_{\pi \gamma}}{dx_\gamma}\right) \right.  = 1
   - \frac{3}{2}\cdot \frac{m_\pi^3}{\alpha} \cdot
  \frac{x_\gamma^2}{1-x_\gamma} \alpha_\pi.
  \label{eq:rxg}
\end{equation}
The measured ratio $R_{\pi}$ if presented in Fig. \ref{fig:R}. The obtained result is 
\begin{equation}
\alpha_\pi\ =\ (2.0\ \pm\ 0.6_{\mbox{\scriptsize stat}}
\ \pm\ 0.7_{\mbox{\scriptsize syst}}) \times 10^{-4}\,\mbox{fm}^3 ~\cite{main} ,
\label{eq:pol2009stat}
\end{equation}
 compatible with the expectation from ChPT. Combination of the COMPASS result with the results of the previous dedicated measurements is presented at the ideogram in Fig \ref{fig:ideo}.
The details of the analysis can be found at \cite{main}.

Nevertheless the uncertainty of the presented  result is still by the factor of two larger than the accuracy of ChPT prediction. New long data taking for pion polarisability measurement was performed at COMPASS in 2012. Analysis of these data is still ongoing. As the result, the improved accuracy of  $\alpha_{\pi}$ measurement under assumption $\alpha_{\pi}+\beta_{\pi}=0$ will be achieved.  In addition the values of $\alpha_{\pi}-\beta_{\pi}$ and $\alpha_{\pi}+\beta_{\pi}$ will be measured independently. Quadrupole polarisabilities of the pion also will be probed. The first result for the polarisability of the charged kaon $\alpha_K$ (since the COMPASS hadron beam has about of 2.4\% of kaons) under assumption $\alpha_{K}+\beta_{K}=const$ is also expected.

\section{Summary}
The measurement of the charged pion polarisability by COMPASS is the most precise dedicated measurement of this value at the moment. The obtained value is compatible with the prediction of  ChPT and  at variance with previous experiments.  This result constitutes important progress towards resolving one of the long-standing issues in low energy QCD. New experimental results for pion and kaon polarisabilities basing on the data collected in 2012 are expected from COMPASS.

\end{document}